\documentclass{bmcart}

\usepackage{amsthm,amsmath}
\usepackage{amssymb}
\RequirePackage{hyperref}
\usepackage[utf8]{inputenc} 
\usepackage{graphicx}

\startlocaldefs

\newcommand{\moneym}[2]{{M}_{#1#2}}
\newcommand{\moneymst}[2]{{M}^{*}_{#1#2}}

\newcommand{\scur}{\mbox{\scriptsize\textit{\textcent}}}
\newcommand{\cur}{\mbox{\textit{\textcent}}}

\newcommand{\blue}[1]{\textcolor{blue}{\bf#1}}
\newcommand{\gold}[1]{\textcolor{orange}{\bf#1}}
\newcommand{\red}[1]{\textcolor{red}{\bf#1}}

\endlocaldefs

\begin{document}

\begin{frontmatter}

\begin{fmbox}
\dochead{Research}


\title{Prospects of BRICS currency dominance in international trade}


\author[
  addressref={aff1},                   
  email={celestin.coquide@utinam.cnrs.fr}   
]{\inits{C.}\fnm{Célestin} \snm{Coquidé}}
\author[
  addressref={aff1},
  corref={aff1},
  email={jose.lages@utinam.cnrs.fr}
]{\inits{J.}\fnm{José} \snm{Lages}}
\author[
addressref={aff2},
email={dima@irsamc.ups-tlse.fr}
]{\inits{D. L.}\fnm{Dima L.} \snm{Shepelyansky}}


\address[id=aff1]{
  \orgdiv{\'Equipe de Physique Th\'eorique,
  	Institut UTINAM, Universit\'e de Franche-Comt\'e,
  	CNRS, 25000, Besançon, France}             
}
\address[id=aff2]{%
  \orgdiv{Laboratoire de Physique Th\'eorique, 
  	Universit\'e de Toulouse, CNRS, UPS, 31062 Toulouse, France}
}



\end{fmbox}


\begin{abstractbox}

\begin{abstract} 
%
During his state visit to China in April 2023, Brazilian President Lula
proposed the creation of a trade currency supported by the BRICS countries.
Using the United Nations Comtrade database, providing the frame of the world trade network associated to 194 UN countries during the decade 2010 - 2020, we study
a mathematical model of influence battle of three currencies, namely, the US dollar, the
euro, and such a hypothetical BRICS currency.
In this model, a country trade preference for one of the three currencies
is determined by a multiplicative factor based on trade flows between countries and their
relative weights in the global international trade. The three currency seed groups are formed by
9 eurozone countries for the euro, 5 Anglo-Saxon countries for the US dollar and the 5 BRICS countries for the new proposed currency.
The countries belonging to these 3 currency seed groups trade only with their own associated currency whereas the other countries
choose their preferred trade currency as a function of the trade relations with their commercial partners. The trade currency preferences of countries are determined on the basis of a Monte Carlo
modeling of Ising type interactions in magnetic spin systems commonly used to model opinion formation in social networks. We adapt here these models to the world trade network analysis.
The results obtained from our mathematical modeling of the structure of the global trade network show that as early as 2012 about 58 percent of countries would have preferred to trade with the BRICS currency, 23 percent with the euro and 19 percent with the US dollar.
Our results announce favorable prospects for a dominance of the BRICS currency in international trade, if only trade relations are taken into account, whereas political and other aspects are neglected.
\end{abstract}


\begin{keyword}
\kwd{World trade network}
\kwd{International trade}
\kwd{Currency}
\kwd{Opinion formation model}
\end{keyword}


\end{abstractbox}
%

\end{frontmatter}




\section{Introduction}

Starting from the Bretton Woods agreements in 1944,
the US dollar (USD) was keeping its dominant position in international trade \cite{wiki1}. Naturally, the  United Nations (UN) reports world
trade transactions between countries in USD \cite{comtrade}.
However, in the last years, a clear tendency emerged to perform trade between certain firms or between certain countries in other currencies than USD.
Thus, Saudi Arabia considers using Chinese yuan (CNY) instead of USD
for the oil sales to China \cite{wallstrj}. There are also other multiple indications
that the USD dominance in the world trade is decreasing
(see e.g. discussions in \cite{FP,BS,GT,balance,bloomberg,liu22,nikkei}).
As an example, CNY becomes the most traded foreign currency on the Moscow Exchange
and surpasses USD \cite{GT,NA}.
In addition, recently, Brazil and China have allowed themselves to carry out commercial and financial transactions directly in CNY or in Brazilian reais (BRL) without resorting to a conversion into USD \cite{fox23}.

In such an atmosphere of de-dollarization of international trade, Brazilian President Lula's call, made during his official visit to China in April 2023, to create a new BRICS currency to end the trade dominance of the dollar has aroused great interest and great concern (see e.g., \cite{FT23,TCP23}).
Thus, it is timely to ask the question of the impact of the creation of such a new BRICS currency on international trade.
Considering this, here we present a mathematical analysis of the possible influence of a new BRICS currency, hereafter referred to as BRI,
using international trade data. More precisely, we develop a model to determine the mathematical preference of a given country to trade in a specific currency which may be different from USD.
Such a mathematical model is built on the World Trade Network (WTN) which is determined from the UN Comtrade database \cite{comtrade} for the period 2010 - 2020.
This database
gives the volumes of monetary transactions between all the countries
during a given year: The money matrix element $M_{c'c}$
gives the total amount of commodities, expressed in
USD of a given year, exported from the country $c$ to the country $c'$.
These money matrix elements $M_{c'c}$ can be used to construct
the Markov chain of trade transactions from which
the WTN.

At present the complex network description finds useful applications
in various fields of science  including social networks,
World Wide Web, biological networks, brain networks and others
(see e.g. \cite{dorogovtsev}). The complex network
properties of the WTN have been studied in
\cite{serrano07,fagiolo09,he10,fagiolo10,barigozzi10,chakraborty18,debenedictis11}.
The Google matrix method \cite{meyer,rmp2015} have been  applied to the WTN in \cite{wtn1}
with the PageRank \cite{brin} construction and in
\cite{wtn1,wtn3} with the use of PageRank and CheiRank vectors
which characterize import and export flows, respectively.
It was shown that PageRank and CheiRank probabilities
obtained from the Google matrix
allow to analyze a crisis contagion in the WTN \cite{wtncrisis}.

The analysis of the competition between two or three currencies
in the WTN requires the development of a new approach
compared to previous WTN studies
where all transactions are expressed in USD.
In the case of two currencies competition (e.g., USD and CNY),
the situation is similar to the problem of spin magnetization, e.g. Ising model, or opinion
formation on simple lattices and complex networks.
Indeed, in spin lattice systems, e.g., a spin up surrounded
by spins down has a tendency to turn down,
taking the direction of dominant neighboring spins.
A similar situation appears also in the problem of opinion
formation on simple lattices or complex networks
when there is a competition between two opinions
or two votes for two different parties. Various models
of opinion formation were proposed and investigated
(see e.g. \cite{galam86,liggett99,galam05,watts07,galam08,castellano09,krapivsky10,schmittmann10}).
For directed social networks, it was pointed that
Page\-Rank weight of nodes (or voters)
can play an important role in an opinion formation process
\cite{kandiah12,eom15}.

The opinion formation model approach
based on PageRank probabilities of complex network nodes \cite{kandiah12,eom15}
has been extended and applied in \cite{coquide23} in order to analyze the trade preferences of world countries
to perform transactions in USD or CNY.
A trade currency preference (TCP) for a given country, i.e., whether the country would prefer to trade in one or another currency,
is determined by two multiplicative factors, namely, the
relative trade volume exchanged with its trade partners and
the global weight of these partners in the global WTN.
The results obtained in \cite{coquide23} show that starting from
2016, the majority of countries in the world would prefer to trade in CNY and no longer in USD as it was the case before 2016.
Of course, these results are based solely on the mathematical analysis
of the WTN transactions and do not take into account political relations between countries. Also, according to the results obtained from the two currencies model \cite{coquide23}, the eurozone (EU) countries are in 2019 on the brink of a USD-to-CNY transition of their trade currency preference.
However, EU countries
usually perform their internal trade in euro (EUR) which is also the second most traded currency in the world. Consequently,
this matter of fact should be taken into account.
Alongside, the hypothetical appearance of a new BRICS currency (BRI) implies that the BRICS countries perform trade between them only in BRI.
Thus, we model the situation where the international trade
is based on three currencies, namely USD, EUR and BRI.
We assume that, similarly to the creation of the euro in the eurozone, the composite BRI currency is based on
the currencies of the BRICS countries, i.e., the Brasilian real (BRL), the Russian ruble (RUR),
the Indian rupee (INR), the Chinese yuan (CNY), and the
South African rand (ZAR).

In our mathematical analysis, we determine a
TCP for a given country indicating that this country has a structural advantage to trade 
with other countries in BRI, EUR, or USD.
This characteristic TCP is based on
the trade flows between countries
obtained from UN Comtrade database \cite{comtrade}.
Extending the approach of \cite{coquide23}, we 
assume that the trade between two countries can be performed in one of the
three currencies BRI, EUR, and USD.
We also define three currency seed groups constituted by countries for which the TCP is always the same. The BRICS group is formed by the BRICS countries, i.e., Brazil, Russia, India, China, and South African. The Anglo-Saxon group is formed by Australia, USA, UK, Canada, and New Zealand.
The EU9 group is formed by Austria, Belgium, France, Germany, Italy, Luxemburg, Netherlands,
Portugal, and Spain. The choice of the 9 EU countries follows
the historical and economical analysis \cite{saint18}
which points out their strong inter-relations. Also, the WTN analysis reported
in \cite{loye21} shows a significant strength of this group
in the international trade.
The number of inhabitants of the Anglo-Saxon and the EU9 groups are comparable, $\sim470$ and $\sim300$ millions of inhabitants, respectively, whereas the BRICS group encompasses a significant larger population, $\sim 3.2$ billions of inhabitants. Let us note that the countries belonging to these 3 currency seed groups never change their TCP. Countries of the BRICS group always prefer to trade in BRI, those of the Anglo-Saxon group always in USD, and those of the EU9 group always in EUR.

The article is organized as follows: the next section presents our 3 currencies model of trade preference using the WTN description of the international trade flows obtained from the UN Comtrade database \cite{comtrade}. Then, the following sections are devoted to the results and discussions.

\section{Model description and data sets}

In this study, we propose a mathematical model of currency competition in the context of the World Trade Network. The WTN is a directed network representing trades between world countries and is constructed from the UN Comtrade database \cite{comtrade} which describes imports and exports of about $10^{4}$ products between the world countries and territories. Here, we consider yearly trades between $N = 194$ countries for the period 2010-2020. In the WTN, the $c'\rightarrow c$ link denotes an export from the country $c'$ to the country $c$ and its weight $M_{cc'}$ is the corresponding exchanged money volume expressed in USD of the considered year. Hence, we note $\moneym{}{c} = \sum_{c'} \moneym{c}{c'}$ ($\moneymst{}{c} = \sum_{c'}\moneym{c'}{c}$) the total import (export) volume associated to the country $c$. Also, we note $M=\sum_cM_c=\sum_cM^*_c$ the total money volume exchanged in the WTN.

The proposed competition model of currencies is an extended version of the model of two currencies studied in \cite{coquide23}, where only USD and CNY were considered. Such a model is similar to opinion formation models applied on social networks to study voting systems coalition formation, strike phenomena (see \cite{galam08,castellano09} for reviews) and on numeric social network such that Twitter \cite{kandiah12}. Here, we propose a model which takes account of 3 currencies, namely, USD, EUR, and BRI. The TCP of a given country depends, both, on the TCP of the other countries and on the probability to import and export with its partners. We consider two stochastic matrices, $S$ and $S^{*}$, encoding import and export trade probabilities between all the countries which constitute the WTN. The matrix element $S_{cc'}=\moneym{c}{c'}/\moneymst{}{c'}$  ($S^{*}_{cc'}=\moneym{c'}{c}/\moneym{}{c'}$) gives the ability of the country $c'$ to export to (to import from) the country $c$.
Also, we define the global import and export trade ability of a country as $P_{c}=\moneym{}{c}/\moneym{}{}$ and $P^{*}_{c}=\moneymst{}{c}/\moneym{}{}$, respectively. The TCP of the country $c$ is a ternary variable $\cur$ which takes the values
$\cur=\mbox{USD}$,
$\cur=\mbox{EUR}$, and
$\cur=\mbox{BRI}$.
The $\cur$ values of the countries belonging to the EU9 group, the Anglo-Saxon group and the BRICS group are kept fixed all along the simulation. For the rest of the world, the TCPs, i.e., either EUR, USD or BRI, are initially randomly affected to the other countries. Hence, the fraction $f_i^{\cur}$ of the world countries initially possess a TCP $\cur$ with $f_i^{\mbox{\tiny USD}}+f_i^{\mbox{\tiny EUR}}+f_i^{\mbox{\tiny BRI}}=1$.
This initial preparation constitute the step $\tau=0$ of the Monte Carlo process. Then, we successively pick at random each one of the $N$ countries for which we compute the following three trade currency scores
\begin{equation}
	Z_{\scur} = \frac{\sum_{c'\neq c} \delta_{\scur,\scur'}(S_{c'c}+S^{*}_{c'c})(P_{c'}+P^{*}_{c'})}{\sum_{c'\neq c} (S_{c'c}+S^{*}_{c'c})(P_{c'}+P^{*}_{c'})},
	\label{eq:dyewtn}
\end{equation}
one for each currency $\cur=\mbox{USD, EUR}$, and $\mbox{BRI}$. In the above equation the sum runs over all the countries $c'$ excepting the country $c$ for which we compute the quantity $Z_{\scur}$, the symbol $\cur'$ stands for the TCP of the country $c'$, and the Kronecker symbol $\delta_{\scur,\scur'}$ is equal to $1$ if $\cur=\cur'$, and $0$ otherwise.
The denominator ensures that the sum of the trade currency scores is equal to 1, i.e., $Z_{\mbox{\tiny USD}}+Z_{\mbox{\tiny EUR}}+Z_{\mbox{\tiny BRI}}=1$.
The $S_{c'c}+S^{*}_{c'c}$ factor encodes the relative commercial strength between the country $c$ and its direct partner $c'$. The factor $P_{c'}+P^{*}_{c'}$ encodes the global trade capacity of the commercial partner $c'$. Accordingly to the values of the three newly computed quantities, $Z_{\mbox{\tiny USD}}$, $Z_{\mbox{\tiny EUR}}$, and $Z_{\mbox{\tiny BRI}}$, the country $c$ TCP possibly changes as it takes the value $\cur$ such as $Z_{\cur}$ is the maximal value of the three. Otherwise stated, the TCP associated to country $c$ becomes, e.g., $\cur=\mbox{EUR}$, if $Z_{\mbox{\tiny EUR}}>Z_{\mbox{\tiny USD}}$ and
$Z_{\mbox{\tiny EUR}}>Z_{\mbox{\tiny BRI}}$. The step $\tau=1$ of the Monte-Carlo process ends once all the countries have been successively picked and, consequently, may have changed their TCPs. The following $\tau$ steps of the Monte-Carlo process reproduce the step $\tau=1$ until a steady state is reached. The average final fraction of countries $f^{\scur}_{f}$ with a TCP $\cur$ is obtained from $10^{4}$ Monte-Carlo simulations. Each one of these simulations starts with a random initial distribution of TCPs, i.e.,
$f^{\mbox{\tiny USD}}_i$, $f^{\mbox{\tiny EUR}}_i$, and $f^{\mbox{\tiny BRI}}_i$.

In the appendix, Fig.~\ref{figS1} illustrates the convergence of the simulation. We observe that the fraction of countries with a given TCP converges rapidly after few Monte-Carlo process steps $\tau\simeq2-3$. We checked that whatever are the initial fractions $f^{\mbox{\tiny USD}}_i$, $f^{\mbox{\tiny EUR}}_i$, and $f^{\mbox{\tiny BRI}}_i$, on average the system reaches always the same steady state characterized by final TCP fractions $f^{\mbox{\tiny USD}}_f$, $f^{\mbox{\tiny EUR}}_f$, and $f^{\mbox{\tiny BRI}}_f$. As an example, we obtain
$f_f^{\mbox{\tiny USD}}=0.24$,
$f_f^{\mbox{\tiny EUR}}=0.27$, and
$f_f^{\mbox{\tiny BRI}}=0.49$, in 2010,
and
$f_f^{\mbox{\tiny USD}}=0.19$,
$f_f^{\mbox{\tiny EUR}}=0.22$, and
$f_f^{\mbox{\tiny BRI}}=0.59$, in 2019.

\section{Results}

Here we present the results obtained from the WTN analysis
with the methods described in the previous Section.

\begin{figure}[th!]
	\begin{center}
		\includegraphics[width=0.98\textwidth]{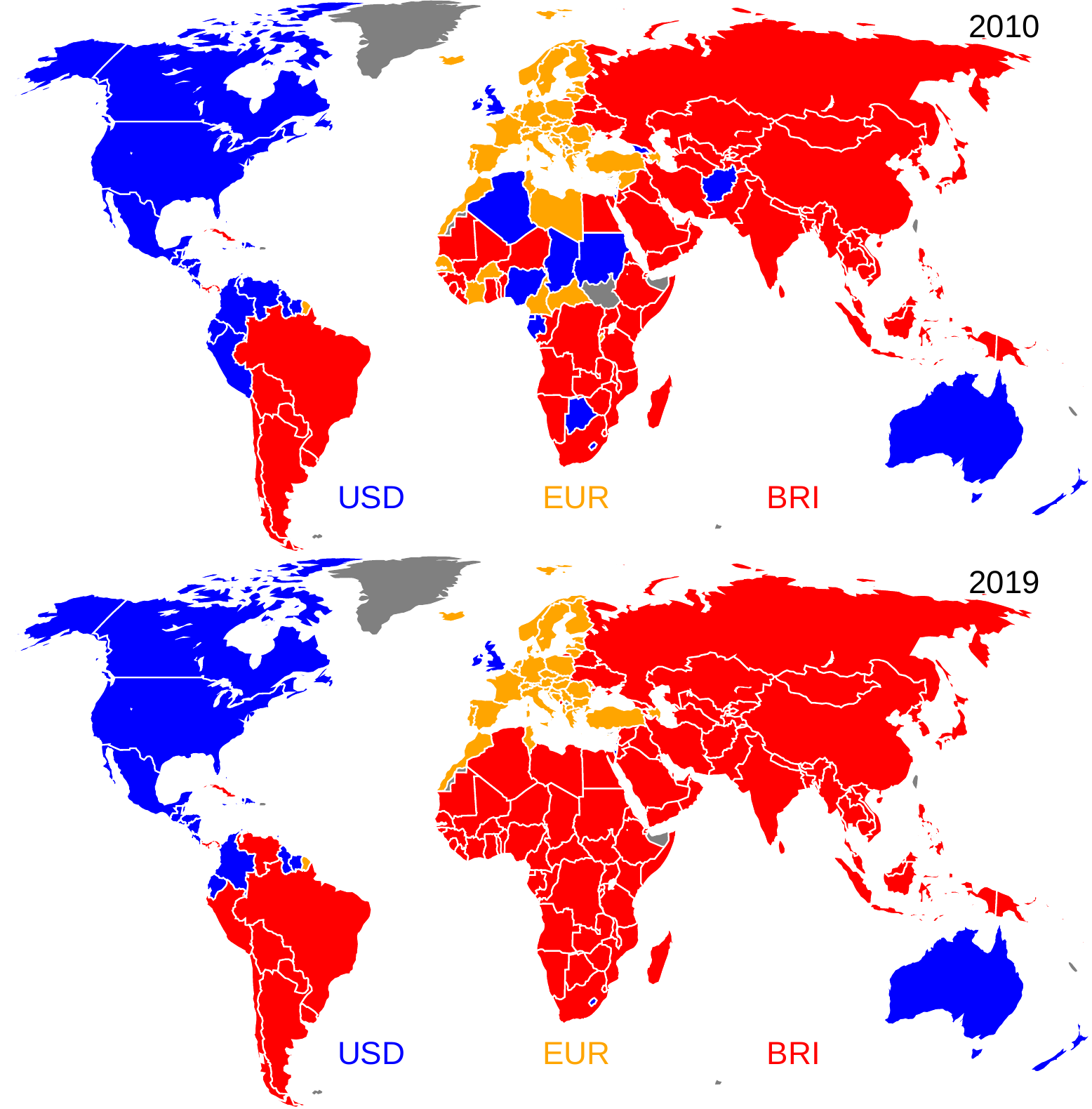}
	\end{center}
	\caption{\label{fig1}\csentence{World distribution of the trade currency preferences, TCP, for 2010 (top panel) and 2019 (bottom panel).} Countries with a trade preference for USD are colored in blue, for EUR in gold, and for BRI in red. Countries colored in grey have no trade data reported in the UN Comtrade database \cite{comtrade}.}
\end{figure}

The world map distributions of TCPs, obtained from the Monte Carlo simulations
based on the calculation of the trade currency scores (\ref{eq:dyewtn}), are shown in Fig.~\ref{fig1} for years 2010 and 2019.
In 2010, the USD TCP is mainly localized in North and Central America (excepting, Cuba and Panama which have a BRI TCP), northern South America, UK, Australia and New Zealand. The EUR TCP is mainly localized in European countries and in countries all around the Mediterranean sea with the exception of Israel and Algeria which have an USD TCP and Egypt which have a BRI TCP. The BRI TCP is located in Asian countries (excepting Afghanistan which have a USD TCP), the countries of the former Soviet Union (excepting Georgia and Azerbaijan which have a USD and EUR TCP, respectively), and the countries of South America (with the exception of the northern countries which have a USD TCP). Apart from the cited exceptions, the distribution of TCPs is quite natural. Indeed, the Americas are divided into a USD block, driven by USA and Canada, and a BRI block driven by Brazil. In Europe, the EUR TCP dominates since the EUR is the official currency of more than 20 countries of the eurozone. Finally, in Asia, the BRI TCP dominates under the influence of China, Russia and India.
By contrast, Africa appears fragmented as the 3 TCPs are comparably distributed all over its countries. This image echoes the post-1989 era battle of influence on African affairs of countries such as France, USA, Russia and China. Globally, the countries of the Southern and Eastern Africa have a BRI TCP, whereas the three types of TCP are quite homogeneously distributed in the Northern, Western and Central Africa. 

From 2010 to 2019, the  EUR group looses
7 countries in Africa but otherwise stays unchanged, and its influence stays focused on the European continent.
From 2010 to 2019, the USD group looses its influence  completely in
Africa (with the exception of the Lesotho) and more mildly in South America where Venezuela and Peru have now a BRI TCP.
On the other hand, from 2010 to 2019, the BRI group has spread over almost the entire African continent and has strengthened its influence in northern South America. In 2019, the BRI influence spans mostly over the developing and least developed countries \cite{dev} and the USD and EUR influences concern mostly the Western world \cite{west}.

By construction, New Zealand and Australia, belonging to the seed countries of the USD group, have always a fixed USD TCP. However, in the 2 currencies model \cite{coquide23}, where only USA and China always trade in USD and CNY, respectively, New Zealand and Australia always have a trade preference for CNY instead of USD.
We note also close similarities between the EUR group (see Fig.~\ref{fig1}) and the swing group observed for 2019 in \cite{coquide23} (see Fig.~4 therein). The swing group in \cite{coquide23} consists in a set of countries which depending on the initial distribution of the TCPs aggregate as a whole to the USD or CNY group. This swing group \cite{coquide23} corresponds to the EUR group presented in Fig.~\ref{fig1} (with the addition of Algeria, Egypt, Ivory Cost, Israel and Jordan).

In the appendix, for the sake of completeness, the world distributions of the trade currency preferences for the years 2012, 2014, 2016, 2018 and 2020 are presented in Fig.~\ref{figS2} which shows indeed a progressive expansion of the BRI trade currency preference over the world.
Also, Tables~\ref{tab:tabS1}, \ref{tab:tabS2}, and \ref{tab:tabS3} give the countries in the USD, EUR, and BRI groups in 2019.

\begin{figure}[th!]
	\begin{center}
		\includegraphics[width=0.8\textwidth]{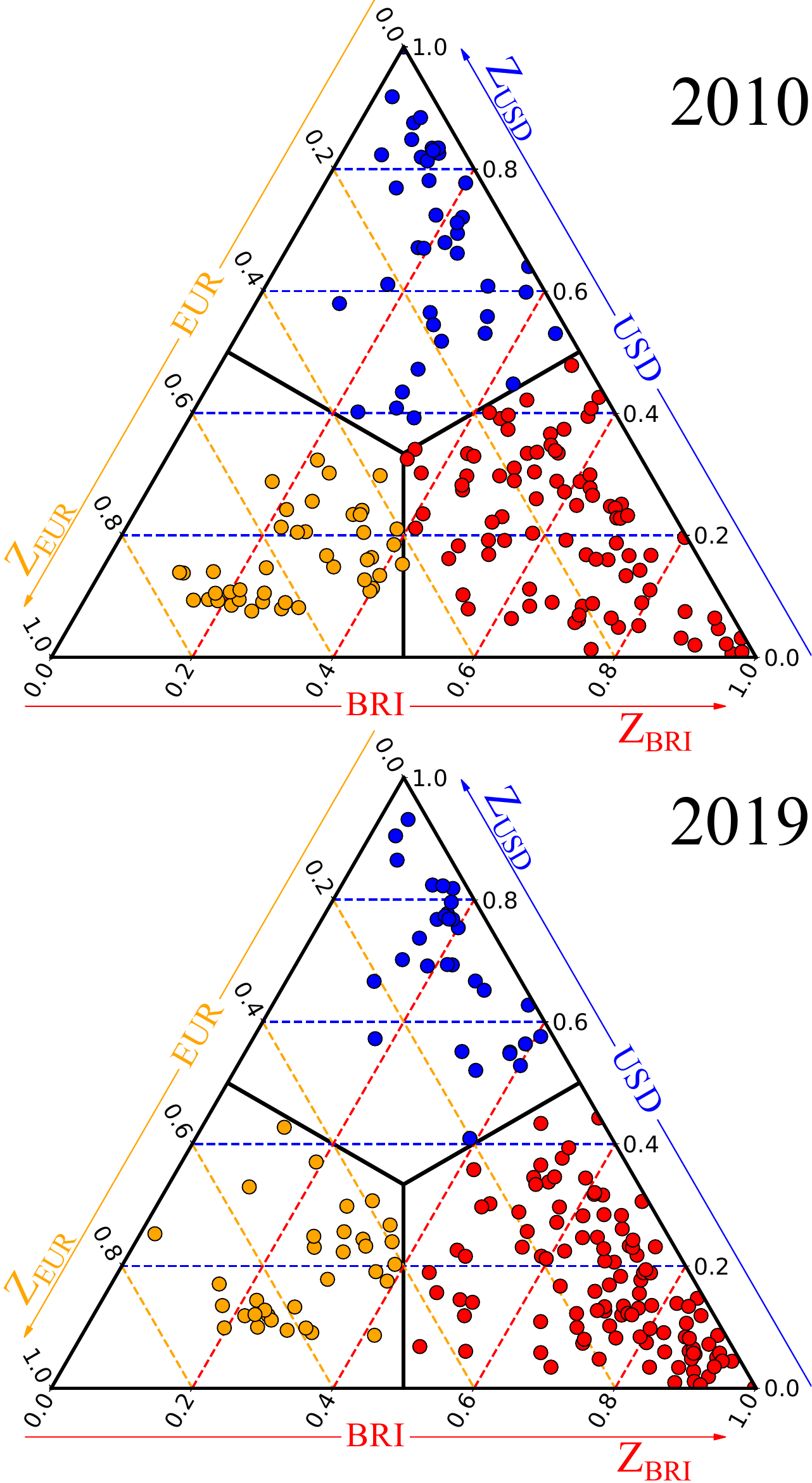}
	\end{center}
	\caption{\label{fig2}\csentence{Distribution of the countries' trade currency ternary scores $\left(Z_{\mbox{\tiny USD}},Z_{\mbox{\tiny EUR}},Z_{\mbox{\tiny BRI}}\right)$ for 2010 (top panel) and 2019 (bottom panel).} A country is represented by a circle. Colors are associated to TCPs, blue for USD, gold for EUR, and red for BRI. The $Z_{\mbox{\tiny USD}}$ coordinate is read along the dashed blue horizontal lines, the $Z_{\mbox{\tiny EUR}}$ coordinate along the gold dashed oblique lines, and the $Z_{\mbox{\tiny BRI}}$ coordinate along the red dashed oblique lines.}
\end{figure}

\begin{figure}[th!]
	\begin{center}
		\includegraphics[width=0.98\textwidth]{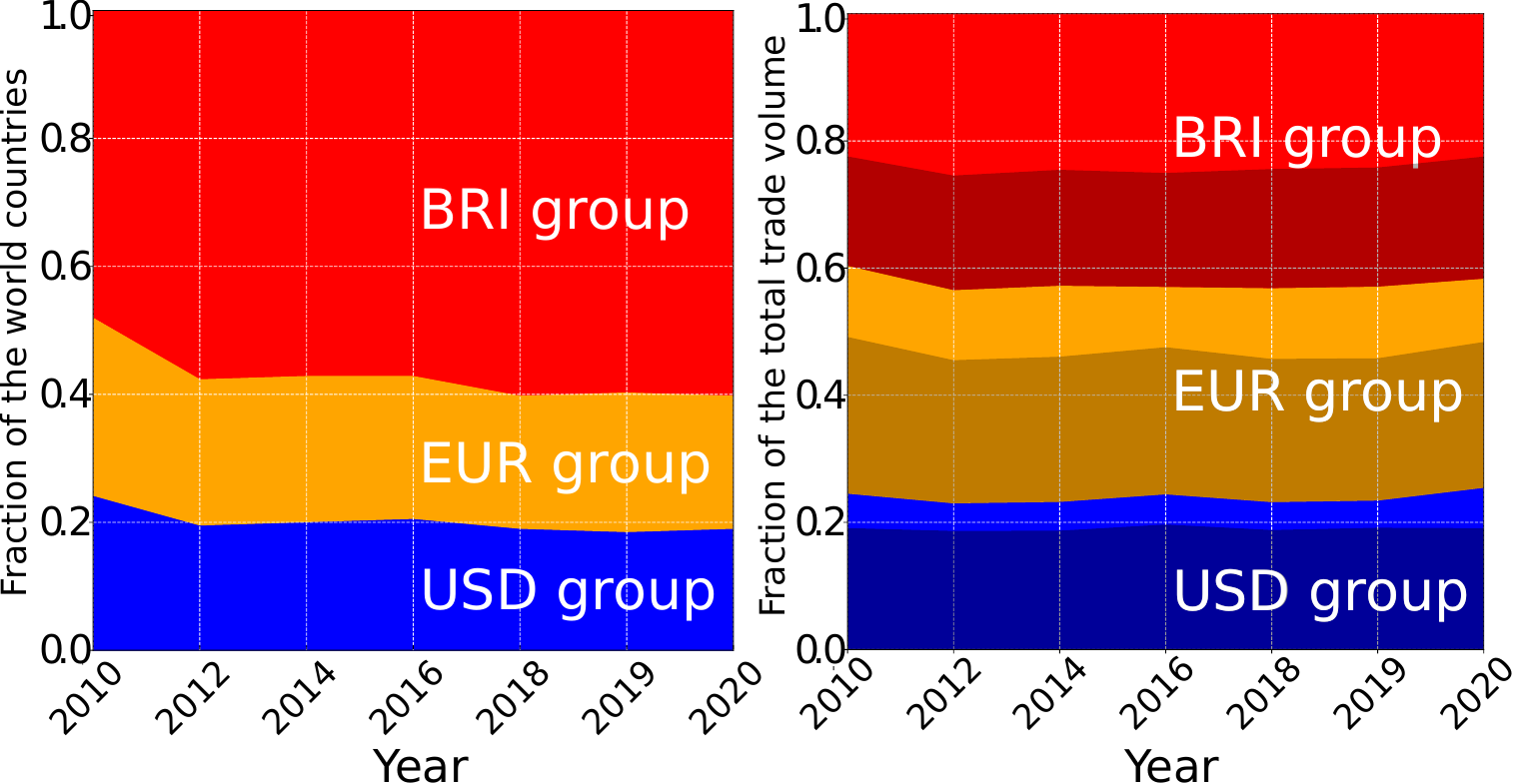}
	\end{center}
	\caption{\label{fig3}\csentence{Evolution of the size of the trade currency groups with time.}
		The height of each band corresponds to the corresponding fraction of world countries (left panel) and to the corresponding fraction of the total trade volume (right panel).
		The group of countries preferring to trade in USD is colored in blue, EUR in gold, and BRI in red. In the right panel, the fractions of the total trade volume associated to the currency seed groups are shown: Anglo-Saxon group (dark blue), the EU9 group (dark gold), BRICS countries (dark red).}
\end{figure}

In Fig.~\ref{fig2}, the distribution of the countries' trade currency ternary scores $\left(Z_{\mbox{\tiny USD}},Z_{\mbox{\tiny EUR}},Z_{\mbox{\tiny BRI}}\right)$ are given for 2010 and 2019.
As in Fig.~\ref{fig1}, for both years, we observe the BRI group gathers more countries than the EUR and USD groups which have similar sizes (see also Fig.~\ref{fig3} hereafter). 
From 2010 to 2019, the EUR group is robust with two main clusters: one cluster located in the range $0.6\lesssim Z_{\mbox{\tiny EUR}}\lesssim0.8$, strongly tied to the EUR currency, which mainly gathers Central Europe and Balkan countries, and the other cluster located in the range $0.4\lesssim Z_{\mbox{\tiny EUR}}\lesssim0.6$ which mainly corresponds to Nordic countries and Baltic countries with the addition of Poland, Switzerland, Greece, Turkey, Azerbaijan and the African countries which are present in the 2010 world distribution of TCPs (see Fig.~\ref{fig1} top panel). From 2010 to 2019, we observe, on average, a shift of these two clusters towards the BRI group. Moreover, most of the African countries present in the EUR group in 2010 moved to the BRI group in 2019, and, e.g., Switzerland and Turkey are in 2019 located close to the $Z_{\mbox{\tiny EUR}}=Z_{\mbox{\tiny USD}}=Z_{\mbox{\tiny BRI}}$ equilibrium point (in the 2020 data, Switzerland has even actually moved into the USD group). Summarizing, a non negligible part of the EUR group countries are on the brink of a transition mainly towards the BRI group.
Also during the 2010-2019 period, the countries of the BRI group moved toward larger values of $Z_{\mbox{\tiny BRI}}$ since in 2019 most of the countries are concentrated in the $Z_{\mbox{\tiny BRI}}>0.6$ zone. This matter of fact indicates a strong entanglement between economies of the BRI group and, mechanically, a weaker dependence on the countries of the EUR and USD groups. More strikingly, we observe in 2019 that most of the USD and BRI groups countries are located in the $Z_{\mbox{\tiny EUR}}<0.2$ zone which suggests that the European countries economies tend to loose their influence on the extra-European economies.

Overall, as shown in Fig.~\ref{figS3} in appendix, the distributions $Z_{\mbox{\tiny USD}}$ and $Z_{\mbox{\tiny EUR}}$ mainly monotonously decrease with the value of $Z_{\mbox{\tiny USD}}$ and $Z_{\mbox{\tiny EUR}}$, respectively. From 2010 to 2019, the lowest range, i.e., $0<Z_{\mbox{\tiny USD}},Z_{\mbox{\tiny EUR}}<0.2$, for both distributions, has even increased, and we note that, for both years, no country has a $Z_{\tiny EUR}>0.8$. This fact corroborates the global decline in the influence of the EUR and USD currencies. On the contrary, the $Z_{\mbox{\tiny BRI}}$ distribution is more homogeneous all over the interval $[0,1]$ with a median which has moved from around $Z_{\mbox{\tiny BRI}}\simeq0.3$ in 2010 to around $Z_{\mbox{\tiny BRI}}\simeq0.7$ in 2019 indicating an increase of the global influence of the BRI.

We checked that a modification of the centrality metrics, replacing, in the trade currency scores (\ref{eq:dyewtn}), the countries
import-export probabilities $P_c$ and $P_c^*$
by the countries PageRank and CheiRank probabilities obtained from the WTN Google matrix \cite{wtn3,coquide23},
leads practically to the same results (compare Fig.~\ref{figS4} with Fig.~\ref{fig1}, the sole differences concern modest size countries, the most visible difference is Suriname which in 2019 has a BRI TCP with the WTN Google matrix and a USD TCP with the present model).

In Fig.~\ref{fig3}, we show the evolution of the fractions
of countries preferring to trade in USD, EUR, and BRI from 2010 to 2020. In 2020, the BRI group
captures 60\% of the world countries, the EUR group 21\% and the USD group 19\%.
However, the trade volume
of the BRI group countries still remains less than 50\% being 42\%,
with 33\% for the EUR group and 25\% for the USD group.
In fact, the fractions of the international trade volume associated to the USD group and to the EUR group are mainly due to the trade exchanges between developed economies within the Anglo-Saxon group and the EU9 group, respectively (see dark colored bands in Fig.~\ref{fig3} right panel).
Also, more than half of the trade volume associated with the BRI group is generated by non-BRICS countries, underlining the fact that the BRI currency is able to influence much more widely than just the BRICS countries, unlike the EUR and USD groups which have little influence beyond their areas of regional and historical influence.

\section{Discussion}

In this work, we analyzed the competition
of three currencies, BRI, EUR and USD,
within the international trade.
We remind that the BRI currency is supposed to be
a new currency pegged to the BRICS countries and proposed recently
\cite{FT23,TCP23}.
For this analysis,
we constructed the WTN for years 2010-2020
from the UN Comtrade database \cite{comtrade}.
As BRI is supported by BRICS, we assume that a group of Anglo-Saxon countries and the EU9 group support the USD and the EUR currencies, respectively.

Our results obtained from Ising type Monte Carlo extensive numerical simulations show that, the influence of each one of the different currencies is well established in their natural zone of influence.
The EUR influence is mainly located in Europe and around the Mediterranean sea.
The USD influence concerns the Anglo-Saxon countries (North America, UK, Australia and New Zealand) but also contiguous regions as Central America and the northern of South America.
The BRI influence spans over Asia and South America on countries which have strong ties with of Brazilian, Russian, Indian and Chinese economies.
While the African continent is fragmented in 2010 between the influence of the three currencies, in 2019 almost all African countries (with the exception of Morocco and Tunisia which have strong historical and economical ties with France) are come under the influence of the BRI. This transition is already well underway in 2012
(see Fig.~\ref{figS2}).
In 2020, 60\% of the countries have a structural trade preference for BRI, 21\% for EUR, and 19\% for USD.
The zone of BRI influence spans from the extended geographical diagonal, from South America to Bering strait and passing by Africa. It encompasses the vast majority of developing and least developed countries. 
The loss of influence of the USD and the EUR echoes the loss of influence of the economies of the Western world which remain confined to their historical zone of influence. 
Moreover, some eurozone countries and others geographically close to it, and historically linked to the EUR, are on the verge of falling under the influence of the other two currencies with a strong tropism for the BRI.

Based on the mathematical analysis of the trade currency preference of the countries, taking account of solely the structure of the WTN and disregarding finer geopolitical considerations, our results show that the influence of the BRICS countries on international trade is now significant and opens the way to a possible prospect of domination of a currency supported by the BRICS development bank at the expense of other global currencies such as USD and EUR.



\begin{backmatter}

\section*{Acknowledgements}
We thank the UN Statistics Division to grant us a friendly access to the UN Comtrade database.

\section*{Funding}
This research has been partially supported by the grant
NANOX N$^\circ$ ANR-17-EURE-0009 (project MTDINA) in the frame  of the Programme des Investissements d'Avenir, France. This research has also been supported by the
Programme Investissements d’Avenir ANR-15-IDEX-0003.

\section*{Abbreviations}
\label{sec:abbreviation}
WTN: World trade network;
CTP: Currency trade preference;
BRICS: Brazil, Russia, India, China, South Africa;
UN: United Nations;
USD: US Dollar;
CNY: Chinese yuan;
BRL: Brazilian real;
EUR: euro;
BRI: BRICS currency;
EU: eurozone;
RUR: Russian ruble;
INR: Indian rupee;
ZAR: South African rand.

ISO 3166-1 alpha-2 code for countries:

AF: Afghanistan; AL: Albania; DZ: Algeria; AS: American Samoa; AD: Andorra; AO: Angola; AI: Anguilla; AQ: Antarctica; AG: Antigua and Barbuda; AR: Argentina; AM: Armenia; AW: Aruba; AU: Australia; AT: Austria; AZ: Azerbaijan; BS: The Bahamas; BH: Bahrain; BD: Bangladesh; BB: Barbados; BY: Belarus; BE: Belgium; BZ: Belize; BJ: Benin; BM: Bermuda; BT: Bhutan; BO: Bolivia; BA: Bosnia and Herzegovina; BW: Botswana; BV: Bouvet Island; IO: British Indian Ocean Territory; VG: British Virgin Islands; BR: Brazil; BN: Brunei; BG: Bulgaria; BF: Burkina Faso; BI: Burundi; KH: Cambodia; CM: Cameroon; CA: Canada; CV: Cape Verde; KY: Cayman Islands; CF: Central African Republic; TD: Chad; CL: Chile; CN: China; CX: Christmas Island; CC: Cocos (Keeling) Islands; CO: Colombia; KM: Comoros; CG: Republic of the Congo; CK: Cook Islands; CR: Costa Rica; CI: Ivory Coast; HR: Croatia; CU: Cuba; CY: Cyprus; CZ: Czech Republic; KP: North Korea; CD: Democratic Republic of the Congo; DK: Denmark; DJ: Djibouti; DM: Dominica; DO: Dominican Republic; EC: Ecuador; EG: Egypt; SV: El Salvador; GQ: Equatorial Guinea; ER: Eritrea; EE: Estonia; ET: Ethiopia; FO: Faroe Islands; FK: Falkland Islands; FJ: Fiji; FI: Finland; FR: France; PF: French Polynesia; FM: Micronesia; GA: Gabon; GM: The Gambia; GE: Georgia; DE: Germany; GH: Ghana; GI: Gibraltar; GR: Greece; GL: Greenland; GD: Grenada; GU: Guam; GT: Guatemala; GN: Guinea; GW: Guinea-Bissau; GY: Guyana; HT: Haiti; HM: Heard Island and McDonald Islands; VA: Vatican; HN: Honduras; HU: Hungary; IS: Iceland; IN: India; ID: Indonesia; IR: Iran; IQ: Iraq; IE: Ireland; IL: Israel; IT: Italy; JM: Jamaica; JP: Japan Ryukyu Island; JO: Jordan; KZ: Kazakhstan; KE: Kenya; KI: Kiribati; KW: Kuwait; KG: Kyrgyzstan; LA: Laos; LV: Latvia; LB: Lebanon; LS: Lesotho; LR: Liberia; LY: Libya; LT: Lithuania; LU: Luxembourg; MG: Madagascar; MW: Malawi; MY: Malaysia; MV: Maldives; ML: Mali; MT: Malta; MH: Marshall Islands; MR: Mauritania; MU: Mauritius; YT: Mayotte; MX: Mexico; MN: Mongolia; ME: Montenegro; MS: Montserrat; MA: Morocco; MZ: Mozambique; MM: Myanmar; MP: Northern Mariana Islands; NA: Namibia; NR: Nauru; NP: Nepal; AN: Netherlands Antilles; NL: Netherlands; NC: New Caledonia; NZ: New Zealand; NI: Nicaragua; NE: Niger; NG: Nigeria; NU: Niue; NF: Norfolk Islands; NO: Norway; PS: State of Palestine; OM: Oman; PK: Pakistan; PW: Palau; PA: Panama; PG: Papua New Guinea; PY: Paraguay; PE: Peru; PH: Philippines; PN: Pitcairn; PL: Poland; PT: Portugal; QA: Qatar; KR: South Korea; MD: Moldova; RO: Romania; RU: Russia; RW: Rwanda; SH: Saint Helena; KN: Saint Kitts and Nevis; LC: Saint Lucia; PM: Saint Pierre and Miquelon; VC: Saint Vincent and the Grenadines; WS: Samoa; SM: San Marino; ST: Sao Tome and Principe; SA: Saudi Arabia; SN: Senegal; RS: Serbia; SC: Seychelles; SL: Sierra Leone; SG: Singapore; SK: Slovakia; SI: Slovenia; SB: Solomon Islands; SO: Somalia; ZA: South Africa; GS: South Georgia and the South Sandwich Islands; ES: Spain; LK: Sri Lanka; SD: Sudan; SR: Suriname; SZ: Swaziland; SE: Sweden; CH: Switzerland; SY: Syria; TJ: Tajikistan; MK: Macedonia; TH: Thailand; TL: Timor-Leste; TG: Togo; TK: Tokelau; TO: Tonga; TT: Trinidad and Tobago; TN: Tunisia; TR: Turkey; TM: Turkmenistan; TC: Turks and Caicos Islands; TV: Tuvalu; UG: Uganda; UA: Ukraine; AE: United Arab Emirates; GB: United Kingdom; TZ: Tanzania; UM: United States Minor Outlying Islands; UY: Uruguay; US: United States; UZ: Uzbekistan; VU: Vanuatu; VE: Venezuela; VN: Vietnam; WF: Wallis and Futuna; EH: Western Sahara; YE: Yemen; ZM: Zambia; ZW: Zimbabwe.

\section*{Availability of data and materials}
The raw data is available from the UN Comtrade database \cite{comtrade}. Additional output data and/or plots of data generated are available upon request.


\section*{Competing interests}
The authors declare that they have no competing interests.


\section*{Authors' contributions}
The authors contributed equally to this work. All authors read and approved the final manuscript.



\bibliographystyle{bmc-mathphys} 
\bibliography{dywtnbib.bib}      


%
%
%
%

\newpage
\section*{Appendix}

\setcounter{table}{0}
\renewcommand{\thetable}{A\arabic{table}}

\begin{table}[h!]
	\centering
	\caption{\label{tab:tabS1}\textbf{List of the 36 countries belonging to the USD group in 2019.} The trade currency preference of these countries is USD at the end of the simulation. The countries are sorted by descending value of $\max\left(P_c,P^*_c\right)$, i.e. the maximum value between the relative import volume $P_c$ and the relative export volume $P_c^*$, and, in case of tie, by descending value of $P^*_c$.
		The bold font countries are the seeds of the USD currency.
		The red colored countries were in the BRI group in 2010. The countries are represented by their ISO2 codes (see \nameref{sec:abbreviation}).}
	\begin{tabular}{|cc|cc|cc|cc|}
		\hline
		\multicolumn{8}{|c|}{USD group countries in 2019}\\\hline\hline
		1.&\textbf{US}&10.&EC&19.&GY&28.&\red{AG}\\
		2.&MX&11.&CR&20.&BS&29.&LC\\
		3.&\textbf{GB}&12.&GT&21.&HT&30.&VC\\
		4.&\textbf{CA}&13.&DO&22.&FJ&31.&KN\\
		5.&\textbf{AU}&14.&HN&23.&SR&32.&WS\\
		6.&IE&15.&SV&24.&BB&33.&DM\\
		7.&IL&16.&NI&25.&LS&34.&GD\\
		8.&CO&17.&TT&26.&BZ&35.&\red{NR}\\
		9.&\textbf{NZ}&18.&JM&27.&CW&36.&\red{TO}\\\hline
	\end{tabular}
\end{table}
\begin{table}[h!]
	\centering
	\caption{\label{tab:tabS2}\textbf{List of the 42 countries belonging to the EUR group in 2019.} The trade currency preference of these countries is EUR at the end of the simulation. The countries are sorted by descending value of $\max\left(P_c,P^*_c\right)$, i.e. the maximum value between the relative import volume $P_c$ and the relative export volume $P_c^*$, and, in case of tie, by descending value of $P^*_c$.
		The bold font countries are the seeds of the EUR currency.
		The blue colored country was in the USD group in 2010. The countries are represented by their ISO2 codes (see \nameref{sec:abbreviation}).}
	\begin{tabular}{|cc|cc|cc|cc|}\hline
		\multicolumn{8}{|c|}{EUR group countries in 2019}\\\hline\hline
		1.&\textbf{DE}&12.&SE&23.&MA&34.&IS\\
		2.&\textbf{FR}&13.&HU&24.&LT&35.&CY\\
		3.&\textbf{NL}&14.&DK&25.&RS&36.&AL\\
		4.&\textbf{IT}&15.&NO&26.&HR&37.&MD\\
		5.&\textbf{BE}&16.&SK&27.&\textbf{LU}&38.&AD\\
		6.&CH&17.&RO&28.&EE&39.&CV\\
		7.&\textbf{ES}&18.&FI&29.&TN&40.&SM\\
		8.&PL&19.&\textbf{PT}&30.&LV&41.&ST\\
		9.&CZ&20.&SI&31.&AZ&42.&\blue{PN}\\
		10.&TR&21.&GR&32.&BA&&\textbf{}\\
		11.&\textbf{AT}&22.&BG&33.&MK&&\textbf{}\\
		\hline
	\end{tabular}
\end{table}
\begin{table}[h!]
	\centering
	\caption{\label{tab:tabS3}\textbf{List of the 116 countries belonging to the BRI group in 2019.} The trade currency preference of these countries is BRI at the end of the simulation. The countries are sorted by descending value of $\max\left(P_c,P^*_c\right)$, i.e. the maximum value between the relative import volume $P_c$ and the relative export volume $P_c^*$, and, in case of tie, by descending value of $P^*_c$.
		The bold font countries are the seeds of the BRI currency.
		The blue (gold) colored countries were in the USD (EUR) group in 2010. The countries are represented by their ISO2 codes (see \nameref{sec:abbreviation}).}
	\begin{tabular}{|cc|cc|cc|cc|}\hline
		\multicolumn{8}{|c|}{BRI group countries in 2019}\\\hline\hline
		1.&\textbf{CN}&30.&PK&59.&MN&88.&\blue{GA}\\
		2.&JP&31.&OM&60.&AM&89.&TJ\\
		3.&KR&32.&KH&61.&BN&90.&SZ\\
		4.&SG&33.&MM&62.&\gold{CM}&91.&\gold{SC}\\
		5.&\textbf{IN}&34.&GH&63.&\blue{VE}&92.&\gold{SY}\\
		6.&VN&35.&UZ&64.&MU&93.&\gold{ME}\\
		7.&\textbf{RU}&36.&LK&65.&ZW&94.&NE\\
		8.&AE&37.&\gold{LY}&66.&ET&95.&SO\\
		9.&MY&38.&BH&67.&\gold{BF}&96.&SL\\
		10.&TH&39.&JO&68.&TG&97.&KP\\
		11.&\textbf{BR}&40.&PY&69.&MR&98.&\blue{GQ}\\
		12.&SA&41.&\gold{CI}&70.&\blue{AF}&99.&MV\\
		13.&ID&42.&UY&71.&KG&100.&\blue{TD}\\
		14.&\textbf{ZA}&43.&MZ&72.&PG&101.&\gold{BI}\\
		15.&PH&44.&KE&73.&MG&102.&BT\\
		16.&CL&45.&CD&74.&BJ&103.&SS\\
		17.&\blue{NG}&46.&BO&75.&GN&104.&GM\\
		18.&UA&47.&AO&76.&ML&105.&DJ\\
		19.&AR&48.&ZM&77.&CU&106.&SB\\
		20.&IQ&49.&NA&78.&YE&107.&TL\\
		21.&BD&50.&PA&79.&UG&108.&ER\\
		22.&KZ&51.&MT&80.&LR&109.&\gold{CF}\\
		23.&\blue{PE}&52.&LA&81.&RW&110.&GW\\
		24.&EG&53.&\gold{SN}&82.&CG&111.&\gold{KM}\\
		25.&KW&54.&TZ&83.&TM&112.&VU\\
		26.&\blue{DZ}&55.&\blue{BW}&84.&MW&113.&\blue{FM}\\
		27.&QA&56.&\blue{GE}&85.&MH&114.&KI\\
		28.&BY&57.&\gold{LB}&86.&\blue{PS}&115.&PW\\
		29.&IR&58.&\blue{SD}&87.&NP&116.&TV\\\hline
	\end{tabular}
\end{table}

\setcounter{figure}{0}    
\renewcommand{\thefigure}{A\arabic{figure}}

\begin{figure}[th!]
	\begin{center}
		\includegraphics[width=0.98\columnwidth]{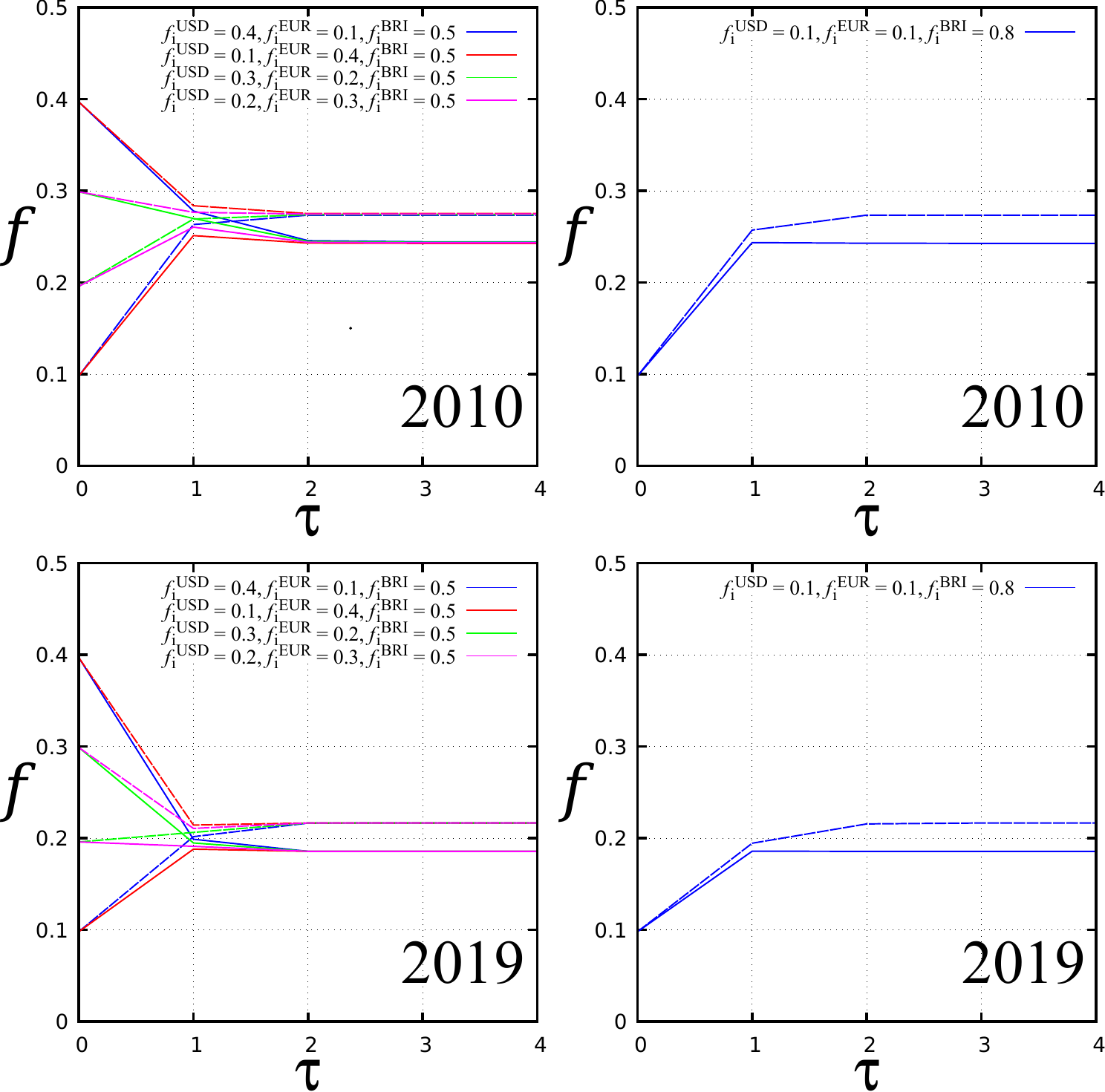}
	\end{center}
	\caption{\label{figS1}\csentence{Examples of evolutions of the fraction $f$ of countries with USD trade preference ($f^{\mbox{\tiny USD}}$, solid line) and EUR trade preference ($f^{\mbox{\tiny EUR}}$, dashed line) as a function of the Monte-Carlo process step $\tau$ in 2010 (top panels) and 2019 (bottom panels).} The complementary fraction $f^{\mbox{\tiny BRI}}=1-f^{\mbox{\tiny USD}}-f^{\mbox{\tiny EUR}}$ gives the fraction of countries with a BRI trade preference.
		The fractions $f$ are averaged over $10^4$ random simulations.
		In left panels, $f_i^{\mbox{\tiny BRI}}=0.5$, we show four configurations such as $f_i^{\mbox{\tiny USD}}+f_i^{\mbox{\tiny EUR}}=0.5$.
		In right panels, $f_i^{\mbox{\tiny BRI}}=0.8$, we show one configuration such as $f_i^{\mbox{\tiny USD}}+f_i^{\mbox{\tiny EUR}}=0.2$. In 2010, the steady state fractions of countries with a USD, EUR, or BRI TCP are $f_f^{\mbox{\tiny USD}}=0.24$, $f_f^{\mbox{\tiny EUR}}=0.27$, and $f_f^{\mbox{\tiny BRI}}=0.49$.
		In 2019, the steady state fractions of countries with a USD, EUR, or BRI TCP are $f_f^{\mbox{\tiny USD}}=0.19$, $f_f^{\mbox{\tiny EUR}}=0.22$, and $f_f^{\mbox{\tiny BRI}}=0.59$.}
\end{figure}

\begin{figure}[th!]
	\begin{center}
		\includegraphics[width=0.98\columnwidth]{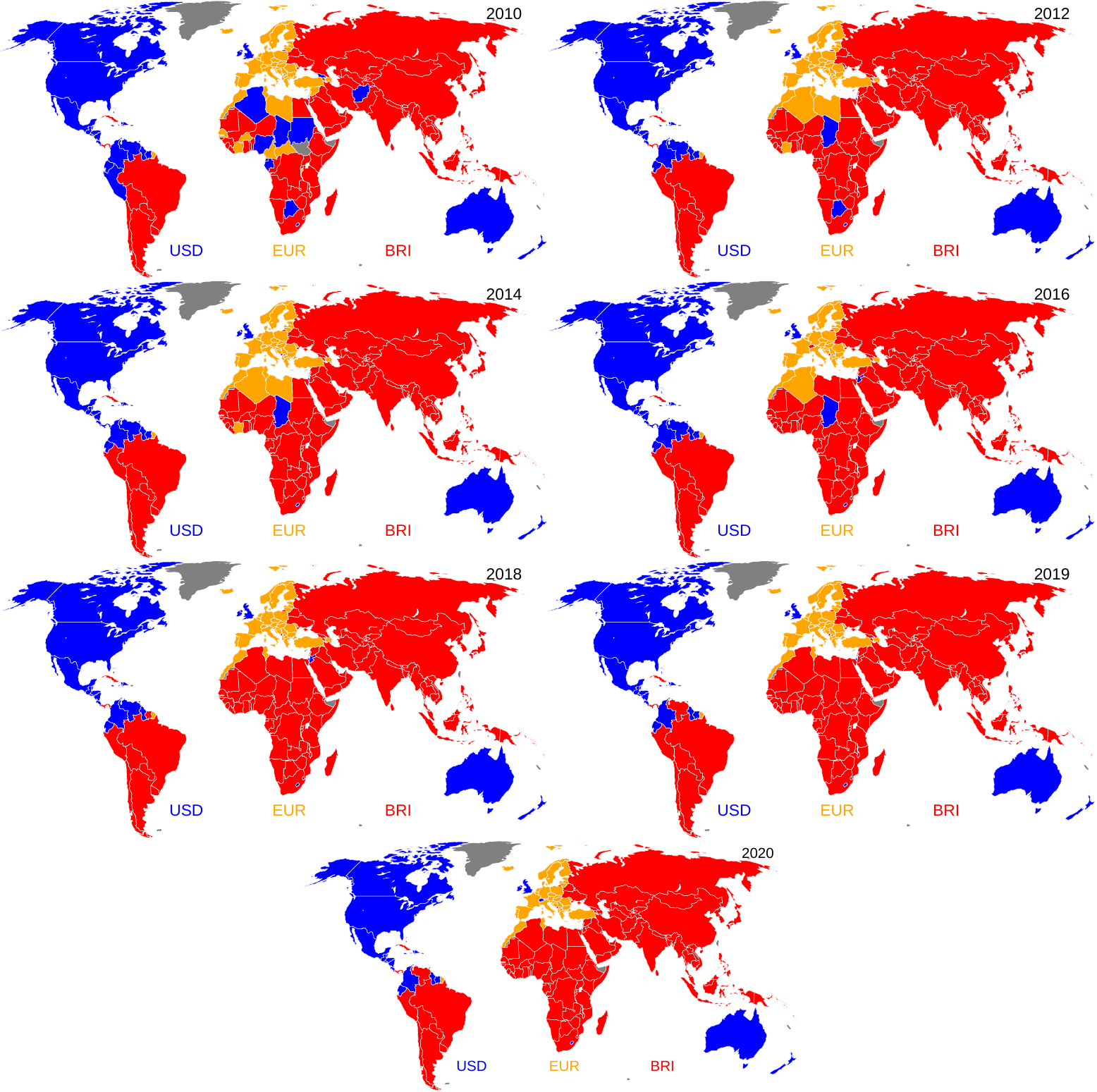}
	\end{center}
	\caption{\label{figS2}\csentence{World distribution of the trade currency preferences for the years 2010, 2012, 2014, 2016, 2018, 2019, and 2020.} Countries with a trade preference for USD are colored in blue, for EUR in gold, and for BRI in red. Countries colored in grey have no trade data reported in the UN Comtrade database \cite{comtrade}.
	}
\end{figure}

\begin{figure}[th!]
	\begin{center}
		\includegraphics[width=0.5\columnwidth]{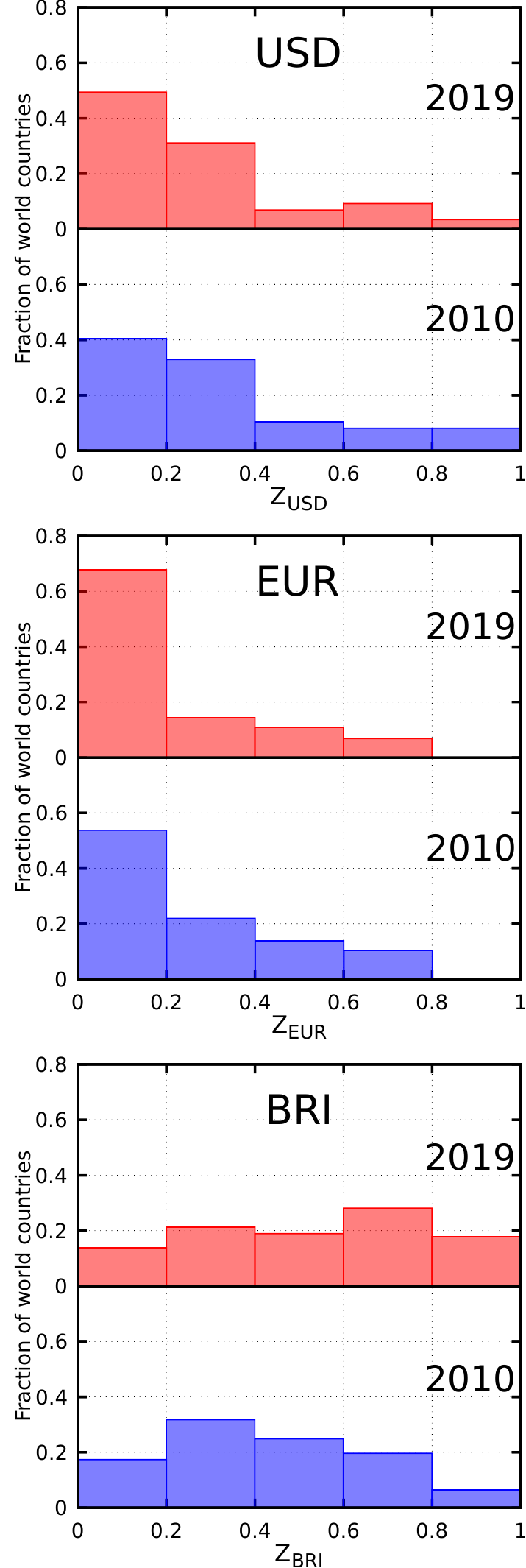}
	\end{center}
	\caption{\label{figS3}\csentence{Distribution of the trade currency scores
		$Z_{\mbox{\tiny USD}}$ (top panel),
		$Z_{\mbox{\tiny EUR}}$ (middle panel),
		and
		$Z_{\mbox{\tiny BRI}}$ (bottom panel) in 2010 (blue boxes) and 2019 (red boxes).} The vertical axis gives the fraction of world countries with $Z_{\scur}$ in a given range.}
\end{figure}

\begin{figure}[th!]
	\begin{center}
		\includegraphics[width=0.98\columnwidth]{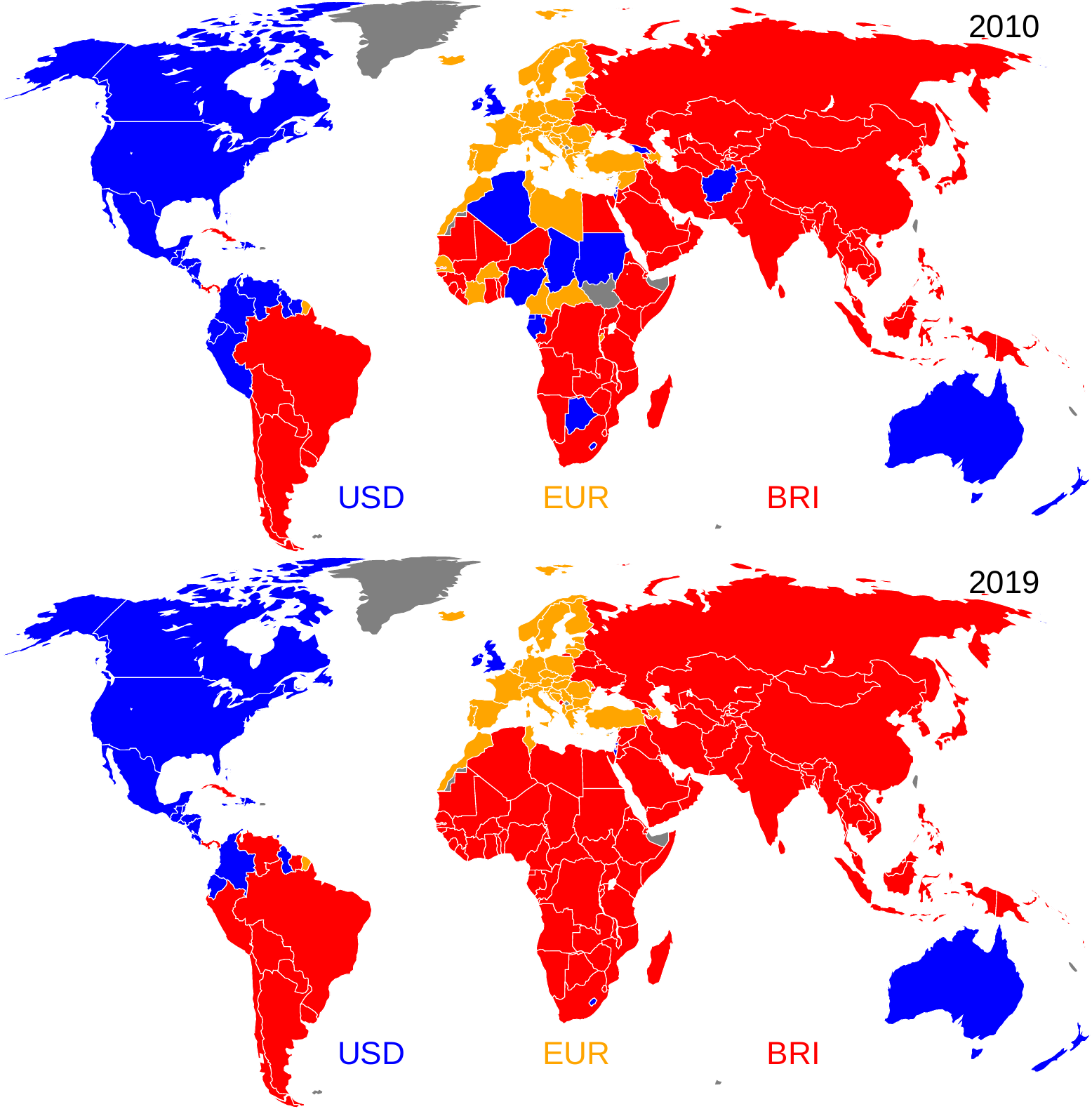}
	\end{center}
	\caption{\label{figS4}\csentence{World distribution of the trade currency preferences for the years 2010 (top panel) and 2019 (bottom panel) when PageRank and CheiRank probabilities, respectively, replace the import and export probabilities, $P_{c'}$ and $P^*_{c'}$, in the trade currency scores $Z_{\scur}$ (\ref{eq:dyewtn}).}
		Countries with a trade preference for USD are colored in blue, for EUR in gold, and for BRI in red. Countries colored in grey have no trade data reported in the UN Comtrade database \cite{comtrade}.}
\end{figure}

\end{backmatter}
\end{document}